\documentclass[aps,twocolumn,showpacs]{revtex4}

\usepackage{graphicx}%
\usepackage{dcolumn,float}
\usepackage{amsmath,amssymb}

\begin{document}

\title{A Density Matrix Renormalization Group study
\\  of Excitons in Dendrimers}


\author{
M.A. Mart\'{\i}n-Delgado$^{\star}$, J. Rodriguez-Laguna$^{\ast}$ and
G. Sierra$^{\ast}$
 }
\address{
$^{\star}$Departamento de
F\'{\i}sica Te\'orica I, Universidad Complutense. 28040 Madrid, Spain.
\\
$^{\ast}$Instituto de Matem\'aticas y F\'{\i}sica Fundamental, C.S.I.C.,
Madrid, Spain. }

\begin{abstract}
We introduce the density matrix renormalization group (DMRG) method as an
efficient computational tool for one-exciton approximations with off-diagonal
disorder. This method allows us to reduce the computational effort
by targetting only a few low-lying eigenstates at each statistical
samplings, in contrast to the exact diagonalization methods that compute
the whole spectrum. As an application of the method, we study excitons in
two families of branched molecules called dendrimers using a recently
introduced simple model.
We compute the absortion peaks for these  dendrimers 
varying their generation number
$g$ and number of wedges $w$.
\end{abstract}


\maketitle

\section{Introduction}
\label{sec1:level1}

\begin{figure}[bp]
\includegraphics[width=5 cm]{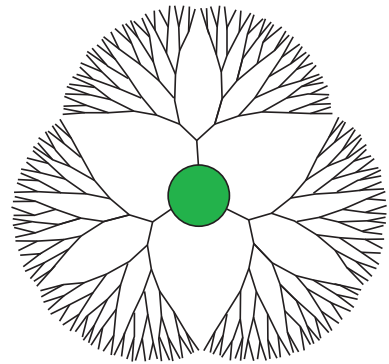}
\caption{A pictorical representation of a dendrimer with parameters
$w=3$, $c=3$, $g=7$.}
\label{fig1}
\end{figure}

\begin{figure}[h]
\includegraphics[width=6 cm]{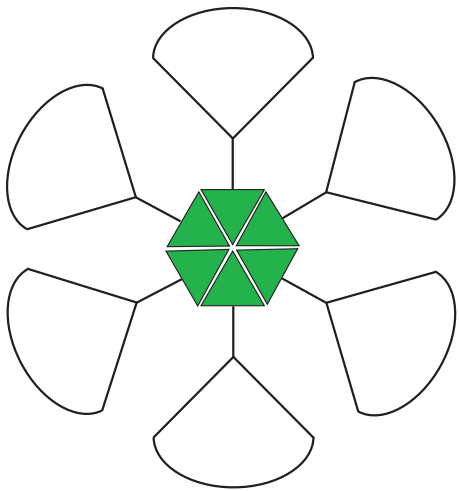}
\caption{Schematical representation of the wedges in a dendrimer with
a core with 6 wedges. The wedges are asembled together as in the convergent
method of sinthesis \cite{frechet}.}
\label{fig2}
\end{figure}

\begin{figure}[H]
\includegraphics[width=8 cm]{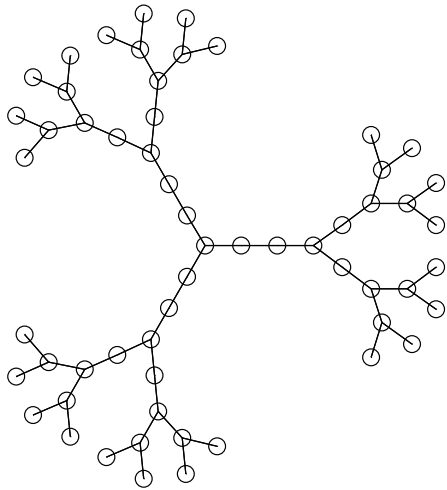}
\caption{An example of extended dendrimer with generation number $g=4$
and wedges $w=3$. At the ramification sites the coordination is $c=3$
like in the compact dendrimers, while the new feature is the presence
of intermediate sites at the legs with $c=2$.}
\label{fig3}
\end{figure}

Dendrimers are large, highly branched polymers \cite{dend} (see
Fig.\ref{fig1}). Some individual dendrimer molecules exceed ten
nanometers in diameter. These hyperbranched molecules are
composed of a central core to which repetitive  dendritic
branches are attached \cite{review0}. Dendrimers have several
distinctive architectural components. Some of them are: i) an
initial core also called {\em focal point}; ii) interior layers
called {\em generations}, composed of repeating units radially
attached one after another. This number determines the branch
length; iii) the degree of {\em connectivity} of each single
molecule in a given site of the macromolecule, i.e., the number
of nearest-neighbours for a given site in the associated lattice
model. This can vary from site to site; iv) the number of initial
strands attached to the focal point can be made variable and
leads to  various separated branches that are called {\em wedges}
or  {\em lobes}. In other words, this is the core multiplicity.
Thus, we can partially characterize the lattice model of a
dendrimer using the notation

\begin{equation}
{\rm D}(w,c,g;N)
\label{1}
\end{equation}

\noindent to denote a dendrimer with $w$ wedges, connectivity
$c$, $g$ generations  and total number of sites $N$. The number of
sites can be obtained from the previous data $(w,g,c)$. In the
following we shall use this nomenclature and a planar
representation of dendrimers (see Figs.\ref{fig2}, \ref{fig3}).
Particular examples of dendrimer lattices are a binary tree, a
Bethe lattice or a Cayley tree.


The field of highly branched macromolecules can also be addressed from the
viewpoint of polymer chemistry. We may classify polymers in two big groups:
linear and branched polymers. Standard polymers elongate in a linear fashion
\cite{review-polymers}, \cite{rva1}, \cite{rva2}, while the new dendritic
polymers are branched \cite{review1}, \cite{review2}, \cite{review3}.
Dendritic polymers differ from linear polymers in that the latter
consists of long chains of molecules, like coils, crisscrossing each
other, while dendrimers  have numerous chain-ends
that can be functionalized. Because of
this, dendritic molecules can be constructed with discrete domains
having different properties.
With a quite descriptive and friendly terminology, standard polymers are called
``spaguetti-like'' while dendrimers are called ``meat-balls''.

The hystorical origins of dendrimers are in the  quest for large,
substrate-selective ligands, in which several research groups
became interested yielding the synthesis of ``tentacle"
\cite{tentacle} and ``octopus" \cite{octopus1,octopus2} molecular
compounds, where long branches radiate from a central hub or a
macrocycle. The first who reported a dendrimer synthesis was
Fritz V\"{o}gtle in 1978 \cite{cascade}, introducing the concept
of cascade reaction and cascade molecules, later known as
dendrimers. Then, research groups led by Tomalia \cite{tomalia}
and Denkewalter \cite{denke} devised routes whereby stepwise
polymerisations could be achieved, producing highly branched
polymers with extremely low polydispersities.

Dendrimers are constructed from branching units and a core.
Their synthesis  begins by attaching
branched molecules to an initial core structure.
Iterative polymerization of more branched
molecules ultimately leads to a globular structure
unable to accommodate further branching because
of steric hindrance.
Using multistep repetitive syntheses, chemists can
construct the dendrimer layer by layer (generation by generation).
This method is known as the ``divergent method."
It was developed  in 1978 by Fritz V\"{o}gtle. It is a difficult and tedious
method for their synthesis,
but it does allow exquisite tailoring of architectural features.
Later, other alternative methods were proposed such as the
``convergent method'',  were dendrimers are built from the surface to the core,
 again using a reiterative sequence of reactions. This method was first introduced in \cite{frechet}.

Dendrimers may have a huge variety of constituent single molecular groups.
Typically they contain  carbon-based organic molecules such as
polyamidoamines, aminoacids, DNA, and sugars,
but also they are being constructed from organosilicons and
organic/inorganic hybrids.

There are many examples of dendrimers whose constitutions have been designed
with a purpose in mind.
By an  accurate choice of the building blocks to be used
and specific functionalization in the periphery (external surface),
many of their properties can be controlled such as
the molecular weight, size, shape,
chirality, density, viscosity, polarity, solubility,
flexibility and surface chemistry of the resulting
macromolecules. This leads to
new materials with a great potential for future applications.

Interest in dendrimers has mushroomed after producing
the wide range of structures  that are known today, and
a great deal of  potential applications for dendrimers have been proposed
\cite{review1,review2,review3}.
Among the many current and future
applications   we may cite drug delivery systems
(in chemotherapy for instance), gene therapy,
materials science, magnetic resonance imaging (MRI) contrast agents,
molecular recognition, antiviral treatment,
energy converters, etc.
In material science, dendrimers may themselves act as
building blocks in the fashioning of larger
supramolecular structures.

Dendrimers include a great volume in relation to their molecular
weight as a consequence of their dendritic structure. They may
quite well be compared to bushes and trees in nature (see
Fig.\ref{fig1}). Cavities inside them can host guest species such
as ions or another molecules. The potential of dendrimers in the
fields of host-guest chemistry and nanotechnology  relies on the
subtle engineering of their architectures to pre-defined designs
and taking advantage of these cavities inside them to encapsulate
the guests.


Of all the many applications of dendrimers, we shall focus on their
optical properties that confer them the potential applications as
light-harvesting molecules, i.e., molecular antennas. There have been
recent experiments  addressing these issues by studying electron transfer
in dendrimers of several types and their optical properties for absortion
and emission of light \cite{kopelman}.

The functioning of dendrimers as single molecule photonic antenna systems
relies on the large number of chain-ends that can be functionalized
to  absorbing photons at the periphery and a subsequent efficient transfer
of the absorbed energy to the center where a fluorescent trap
or a laser dye can be  placed at the
dendritic core \cite{kopelman}.

Following these experimental studies, we shall focus for the
moment in two types of dendrimer familes: {\em compact
dendrimers} and {\em extended dendrimers}. They are shown in
Fig.\ref{fig3}.  These are phenylacetylene dendrimers where at each
node or site of the dendrimer lattice there are benzene molecules.
Depending on the position of these benzenes, the branching of the
site can be either two ({\em para} position) or three ({\em
ortho} position). Compact dendrimers consists of three Cayley
trees joined together at the focal point. This implies that the
number of wedges is $w=3$. All their sites exhibit ortho
positions leading to a connectivity $c=3$. The number of
generations is variable. In \cite{kopelman} dendrimers up to
generation 5 were studied. Thus, a compact dendrimer is a ${\rm
D}(3,3,g;N(g))$-dendrimer (\ref{1}) with 

\begin{equation}
N(g)= 1+3\sum_{n=1}^g 2^{n-1} = 3 \times 2^{g}-2.
\label{1b}
\end{equation}

\noindent This is an
exponential growth. The first elements of this series are D4,
D10, D22, D46, D94, etc. The experimental studies of
\cite{kopelman} show that the electronic excitations of these
fractal structures are of a localized nature.

Extended dendrimers have also a threefold symmetry around the focal point,
but they have generation-dependent segments lengths. Namely,
their sites can be of variable connectivity, either $c=2$ (para position)
or $c=3$ (ortho position) depending on the generation. Thus, an extended
dendrimer is a ${\rm D}(3,[2,3],g;N(g))$-dendrimer (\ref{1}) with

\begin{equation}
\begin{aligned}
N(g) & =1+3\sum_{n=1}^g (g-n) 2^{n-1}+3 \ 2^{g-1} \\  
     & = 9 \times 2^{g-1}-3g-2, 
\end{aligned}
\label{1c}
\end{equation}

\noindent where the factor in (\ref{1c})
$l(g,n)\equiv g-n+$ ($n=1,\ldots,g$)
is the length of each segment in the dendrimer.
This is also an exponential growth.
The first elements of this series are D4, D10, D25, D58, D127, etc.

Despite this small difference in the structure of extended dendrimers,
it has important physical consequences which have been reported experimentally
\cite{kopelman}: electronic excitations become delocalized with increasing
size (generation) of the supermolecule. This experimental evidence comes
from the study of dendrimer absortion spectra, which clearly shows that
the first absortion peak for the compact family remains constant at a given
wave length, while in the extended family this wave length decresase with
increasing size. In other words, in extended dendrimers the gap for excitations
from the ground state to the first excited states tends to close while it
remains constant in compact dendrimers.
Thus, there is experimental evidence for the following scenario:

\begin{equation}
\begin{aligned}
{\rm Compact \; Dendrimers} & \longleftrightarrow  {\rm Gapped  \; Spectrum} \\
{\rm Extended \; Dendrimers} & \longleftrightarrow {\rm Gapless  \; Spectrum}
\end{aligned}
\label{2}
\end{equation}

A complete {\em ab initio} computation of the electronic properties in these
two families of dendrimers would require to include many different effects and
a truly many-body model for these supermolecules. However, recently a very
simple phenomenological model of one-exciton processes
has been proposed by Harigaya
\cite{harigaya1,harigaya2,harigaya3} which produces good fits for the
experimental absortion peaks in extended dendrimers.

In this paper our aim is to propose the density matrix renormalization
group (DMRG) method as a good computational tool for dealing with
one-exciton process in the presence of disorder. This method allows us to
perform an exhaustive study of excitons in the Harigaya model to be explained
below, for both families of dendrimers.

Our studies are theoretical, we consider  dendrimers as a useful model
for investigating the dependence of physical properties on molecular size
and topology. This paper is organized as follows: in Sect. II we introduce 
a simple Frenkel Hamiltonian describing exciton processes and concentrate
in one-exciton approximations using a recently introduced model Hamiltonian;
in Sect. III we propose the DMRG method as an appropriate computation method
for excitons with disorder in extended and compact dendrimers, and we present
our numerical results in a variety of situations with different numbers
of generations $g$ and wedges $w$. Sect. IV is devoted to conclusions and
future prospects.


\section{Excitons in Dendrimers}
\label{sec2:level1}

A consequence of the aforementioned experimental studies  \cite{kopelman}
is that the family of extended dendrimers can serve as
artificial light-harvesting antennas. This has been demonstrated
experimentally \cite{antennas}. The light is harvested over a wide area
on the surface of the dendrimer and funneled by dipole-dipole interactions
to the single active site at the focal point where energy conversion
takes place.
 Based on the optical spectra of the
first absortion peak \cite{kopelman}, Kopelman et al. conjetured
that the electronic excitations in extended dendrimers are localized
on the linear segments of the branched molecules.
Theoretical studies have confirmed this fact showing that the relative
motion of photogenerated electron-hole pairs (excitons)
is confined to the various segments and energy-transfer
may then be described by the Frenkel exciton model \cite{klafter}.

\begin{figure}[h]
\includegraphics[width=7 cm]{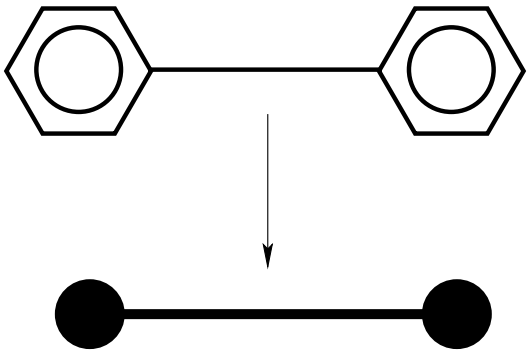}
\caption{Simplified treatment of the phenil rings as single sites.}
\label{fig4}
\end{figure}

In the Frenkel exciton model, at each site of a lattice there is
an active center modeled with a two-level system. At half-filling,
there is an electron per site on average and in the absence of
interactions among centers, the ground state $|{\rm GS}\rangle$ is
formed by putting each electron at the corresponding ground
states $|{\rm vac}\rangle$ of the two-level systems, that is,

\begin{equation}
|{\rm GS}\rangle = |{\rm vac}\rangle_1  |{\rm vac}\rangle_2 \ldots
|{\rm vac}\rangle_N
\label{3}
\end{equation}

A local excitation is formed by promoting one of the electrons at
a given site to the first excited level $|{\rm exc}\rangle$.
This is a localized
electron-hole pair that models an exciton. An example of this is
the state

\begin{equation}
|{\rm Ex}\rangle = |x\rangle =
|{\rm vac}\rangle_1 \ldots |{\rm exc}\rangle_x \ldots |{\rm vac}\rangle_N .
\label{4}
\end{equation}

Let us introduce fermionic creation and annihilation operators
$c^{\dagger}_{i,\alpha}, c_{i,\alpha}$ for these localized centers,
where $i$ denotes the lattice site and $\alpha$ is a label that stands
for the energy levels: $\alpha = 0$ for $|{\rm vac}\rangle$ and $\alpha = 1$
for $|{\rm exc}\rangle$. Then, the localized energy levels can be represented
as

\begin{equation}
\begin{aligned}
|{\rm vac}\rangle_i = c^{\dagger}_{i,0} |0\rangle \\
|{\rm exc}\rangle_i = c^{\dagger}_{i,1} |0\rangle
\label{4a}
\end{aligned}
\end{equation}

\noindent where $|0\rangle$ is the vacuum of the fermionic operators.
The excitonic operator that creates one exciton from the localized
ground state $|{\rm vac}\rangle_i$ is denoted by $b^{\dagger}_i$
and acts as follows

\begin{equation}
|{\rm exc}\rangle_i = b^{\dagger}_i |{\rm vac}\rangle_i, \; {\rm with} \;
b^{\dagger}_i = c^{\dagger}_{i,1} c_{i,0}.
\label{4b}
\end{equation}

\noindent The excitonic operator $b^{\dagger}_i$ creates a localized
electron-hole pair at the site $i$ and has bosonic character.

When interactions
among centers are brought to the system, the interesting question
is to see how one these excitons evolve under quantum
fluctuations. A simple model for these interactions is presented
in the Frenkel Hamiltonian, namely,

\begin{equation}
H = \sum_{i\in \Lambda} E_i \ b^{\dagger}_i  b_i + \sum_{\langle
i,j\rangle} J_{i,j} \ (b^{\dagger}_i b_j + b^{\dagger}_j b_i)
\label{5}
\end{equation}

\noindent
where $E_i$
are chemical potentials that represent the energy gaps at each
site of the lattice $\Lambda$ and $J_{i,j}$ is a hopping integral
for nearest-neighbour centers $\langle i,j\rangle$ that introduce
interactions between the site centers \cite{xy}. The origin of these
couplings is the residual electric dipole interactions between
the molecules modeled by the active centers. It is assumed that
this interactions are short-ranged. Their effect is to allow
excitation transitions between nearest-neighbours sites.

Harigaya has recently proposed to study one-exciton processes in
the family of extended dendrimers introducing off-diagonal
disorder \cite{harigaya1,harigaya2,harigaya3}. This simple model
produces a good accurate phenomenological fit to the experimental
data of the first absortion peak reported in \cite{kopelman}.
Harigaya considers a first simplification of the molecular
estructure in extended dendrimers in order to make the analysis
more tractable. This simplification amounts to replace the
complicated phenil rings by simple activation centers as depicted
in Fig.\ref{fig4}. A second aproximation considered in
\cite{harigaya1,harigaya2,harigaya3} is to restrict the study to
the one-exciton aproximation within the Frenkel Hamiltonian
(\ref{5}) for excitons. We shall refer to this simple model based
on these two aproximations as the Harigaya model. A further
assumption in this model is the presence of off-diagonal disorder.

To derive the one-exciton approximation of the Frenkel Hamiltonian (\ref{5})
we project this Hamiltonian onto the Hilbert space of one-excited states
(\ref{4}) by considering a general superposition of localized excited states
as follows

\begin{equation}
|\Psi_{{\rm ex}}\rangle = \sum_{x\in \Lambda} C_x |x\rangle
\label{6}
\end{equation}

\noindent where $ C_x$ are unknown amplitudes of the one-exciton states.
Next,  we set up the Schr\"{o}dinger equation in this sector

\begin{equation}
H |\Psi_{{\rm ex}}\rangle = {\cal E}_{{\rm ex}} |\Psi_{{\rm ex}}\rangle
\label{7}
\end{equation}

The solution of this equation is equivalent to solving the
Schr\"{o}dinger equation for an associated one-body Hamiltonian
$H_{{\rm 1 ex}}$ that takes the following form

\begin{equation}
H_{{\rm 1 ex}} = \sum_{i\in \Lambda} E_i |i\rangle \langle i| +
\sum_{\langle i,j\rangle} (J_{i,j} |i\rangle \langle j| + {\rm h.c.})
\label{8}
\end{equation}

\noindent where $|i\rangle$ are one-particle states representing
an exciton state at site $i$. That is, an onsite exciton state is
assigned to each phenyl ring (see Fig.\ref{fig3} and
Fig.\ref{fig4}).

In extended dendrimers, interaction strengths between neighouring dipole
moments may vary among position of dipole pairs. In the Harigaya model
it is assumed that they are randomly distributed and the following Gaussian
distribution function is chosen for the coupling strenghts

\begin{equation}
P(J_{i,j}) = {1\over \sqrt{2\pi} J} {\rm e}^{-{1\over 2} ({J_{i,j}\over J})^2}
\label{9}
\end{equation}

\noindent where $J$ is the standard deviation of the interaction and the
mean value of the interaction is taken as zero.

\noindent Further, the excitation energies $E_i$ are also assumed
to be uniformly distributed

\begin{equation}
E_i = E, \ \forall i.
\label{10}
\end{equation}

Thus, the two adjustable parameters of the model are $E$ and $J$. The former
specifies the central energy position of excitons in the optical spectra.
According to \cite{harigaya1,harigaya2,harigaya3}, 
we shall take the values of these parameters as $E=37200 {\rm cm}^{-1}$ 
and $J=3552 {\rm cm}^{-1}$, which were found to give the best fit to the
experiments \cite{kopelman}.

The diagonalization of (\ref{8}) under these assumptions gives the energies of
one exction states measured from the ground state.
The energy position of the optical absortion edge is always given by
the lowest eigenvalue because the state with the lowest energy is
always allowed for dipole transition from the ground state
\cite{harigaya1,harigaya2,harigaya3}. Thus, the values of the absordion
edge $E_{{\rm ab}}$ (energy of the lowest optical excitation)
for the family of extended dedrimers is computed as

\begin{equation}
E_{{\rm ab}} = E - \langle e_0 \rangle J
\label{11}
\end{equation}

\noindent where $\langle e_0 \rangle$
is the average of the ground state energies $e_0$ obtained by
diagonalization and sampling of the one-exciton  Hamiltonian in eq. (\ref{8}).

To solve this model of stochastic fluctuations in the
Hamiltonian matrix elements
that simulates the random distribution of couplings, in
\cite{harigaya1,harigaya2,harigaya3} an exact diagonalization procedure
was employed for a sampling of 10000 realizations. This was carried out
up to the fifth generation of extended dendrimers which has D127 sites.

As in the computation of the first peak of the absortion spectra
only the computation of the ground state of (\ref{8}) matters, then it is
clear that  using  an exact diagonalization method for solving a
$127\times 127$ matrix a number of 10000 times implies a waste of computer
resources. Of all the over hundred eigenvalues, only  one is needed.
Thus, it would be more convenient to have a method that targets
only the desired
state with the same accuracy as the exact diagonalization method while not
having to compute the remaining  states.
In next section we shall compute this absortion peak using the
density matrix renormalization group method \cite{white1}, \cite{white2},
\cite{noackwhite}.

\begin{figure}[h]
\includegraphics[width=8 cm]{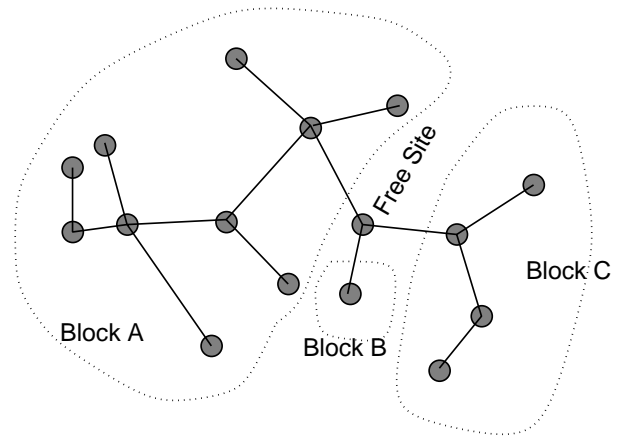}
\caption{Superblock decomposition of the lattice associated to a general
tree graph (no cycles allowed).
The free site denotes the probing site $\bullet$ in the block splitting
${\rm A B C} \bullet$.}
\label{fig5}
\end{figure}

\section{DMRG for Excitons: Results }
\label{sec3:level1}

We propose to use the DMRG method as explained in \cite{qm-dmrg} for
the computation of one-exciton problems with off-diagonal disorder.
This method has been used to compute with high accuracy the low lying
energy states in quantum mechanical problems \cite{delta}, not only in
one dimensional lattices but also in two and three dimensions \cite{prg}.
The advantages of this quantum mechanical DMRG (QM-DMRG) over
exact diagonalization models in one-exciton calculation  are the following:

\begin{enumerate}

\item It is possible to reach lattice sizes that are out
of the reach for exact diagonalization techniques, while keeping the
same degree of accuracy in the determination of energies and wavefunctions.

\item It is possible to save both CPU time and space
for the computations in each sampling. This is specially interesting
when the number of samplings $N_S$ is a large number.

\end{enumerate}

\begin{figure}[h]
\includegraphics[width=8 cm]{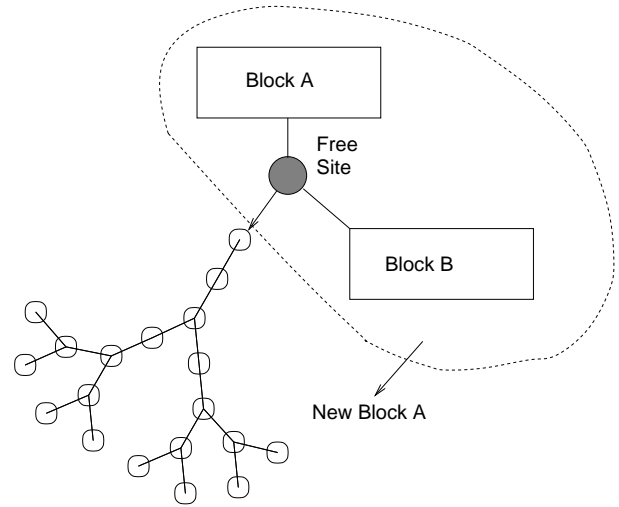}
\caption{Schematic representation of the sweeping process in the DMRG process
for dendrimers. The free site is at a site with coordination number $c=3$.}
\label{fig6}
\end{figure}

These properties make the QM-DMRG method specially well-suited for excitonic
calculations in dimensions higher that one when computation means due
to disorder and connectivity become quite demanding.

We have tested this proposal in the interisting case of extended and
compact dendrimers. With this method we can increase the number of
generations $g$ for each dendrimer family
to sizes not reachible with other methods.

We shall not dwell upon all the details of the QM-DMRG formulation that
can be found in \cite{qm-dmrg,delta,prg},
but instead we shall stress the main peculiarities
of the method when applied to extended and compact dendrimers.

\subsection{DMRG for Dendrimer Lattices}

For the sake of concreteness, we shall make the discussion in terms
of dendrimers with a fixed number of wedges $w=3$ and connectivity
$c=3$, motivated by the Harigaya model, but the method is equally
applicable to a generic tree graph. Later, we shall present results
when the number of wedges is different ($w\neq 3$).

In the standard formulation of the DMRG, the whole lattice also
called {\em universe}  ${\cal U}$ or superblock (SB)
is split into left ($B_L$) and right ($B_R$)
blocks according to the decomposition: ${\cal U} = B_L \bullet B_R$ or
${\cal U} = B_L \bullet \bullet B_R$. The blocks describe the degrees of
freedom of the system and the environment, which can be represented either
by $B_L$ or $B_R$ depending on which  stage of the RG-sweeping process
we are. In the case of dendrimers, we find more appropriate to use the
first decomposition based on one site, ${\cal U} = B_L \bullet B_R$.
The single lattice site  $\bullet$ connects the two blocks and serve as
a {\em probe} to test the reaction of the system degrees of freedom to the
coupling to the rest of the environment.
This probing site is a movable site that runs all over the lattice during
the sweeping process.

\begin{figure}[h]
\includegraphics[width=8 cm, height=8 cm]{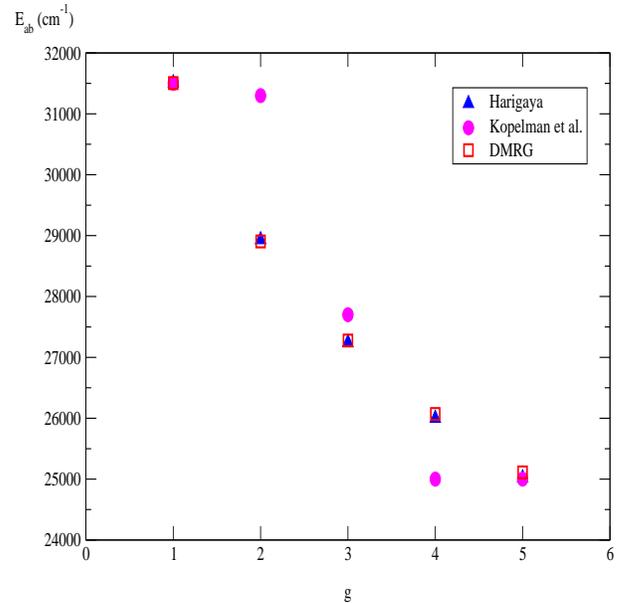}
\caption{Results for the first absortion peak $E_{{\rm ab}}$
as in (\ref{11}) for extended dendrimers versus the
generation number $g$. Two numerical methods are
compared with the experimental results of Kopelman et al.
\cite{kopelman}: DMRG (ours) and ED (Harigaya \cite{harigaya1}).}
\label{fig7}
\end{figure}

Let us introduce the following notation for the Hamiltonian matrix
elements in (\ref{8}):

\begin{equation}
H_{{\bf i},{\bf j}} =
\begin{cases}
E  & \quad {\bf i}={\bf j} \\
J & \quad \parallel {\bf i}-{\bf j} \parallel  = 1 \\
0                   & \quad \text{otherwise};
\end{cases}
\label{12}
\end{equation}

\noindent where ${\bf i}, {\bf j}$ are vectors of integer
components in a dendrimer lattice. We may use a binary notation
for labeling these sites.

Although the planar dendrimer lattice extends all over the plane,
there is a distinctive feature that makes it closer to a one single chain.
Namely, given any two points, there exists a unique path joining them,
unlike a real two-dimensional  lattice like the square lattice.
This characteristic is reminiscent of the left-right directions in
one-dimensional chains and it is the basis for trying an application of
DMRG here \cite{prg}.
Next we briefly describe the several ingredients entering in the application
of DMRG to dendrimer lattices.

\begin{figure}[h]
\includegraphics[width=8 cm, height=8 cm]{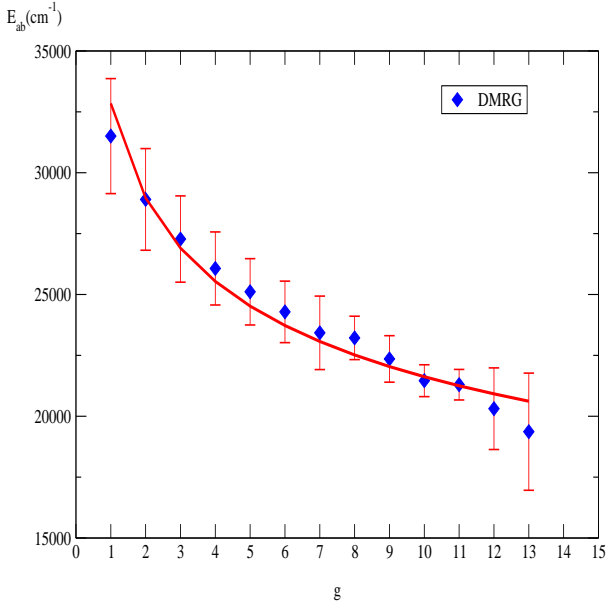}
\caption{DMRG results for the first absortion 
peak $E_{{\rm ab}}$  (\ref{11}) in extended dendrimers versus 
the generation number $g$\cite{sampling}.} 
\label{fig8}
\end{figure}

\subsubsection{Superblock Decomposition of the Dendrimer}

The standard DMRG decomposition of the original lattice to
perform the truncation/renormalization process is ${\cal U} =
B_{l}^L \bullet \bullet B_{N-l-2}^R$ where the subscripts $l$ and
$N-l-2$ denote the number of sites inside each left and right
blocks such that their sum - plus two -  equals the total number
of lattice sites $N$. The index $l$ denotes the iteration step of
the RG process. This decomposition applies both to
one-dimensional lattices as well as higher dimensional cases
\cite{prg}.

In our case, for a generic dendrimer with a number $w$ of wedges
and sites with connectivity $c$,
we use a superblock decomposition as follows

\begin{equation}
{\cal U} = B_{1}(l)B_{2}(l)\ldots B_{c}(l) \bullet_l,
\label{13}
\end{equation}

\noindent where the probing site is joint to the wedge-blocks (see
Fig.\ref{fig5}) and it is a movable point that traverses the
dendrimer lattice as the sweeping process takes places. We denote
it as a free point in Fig.\ref{fig5} and it is a generic point of
the lattice for a given RG-step $l$, not always the focal point
at the core. This decomposition is generic of a tree graph with
no cycles and the DMRG decomposition is applicable with
generality to any graph, as shown in  Fig.\ref{fig5}.

\begin{figure}[h]
\includegraphics[width=8 cm, height=8 cm]{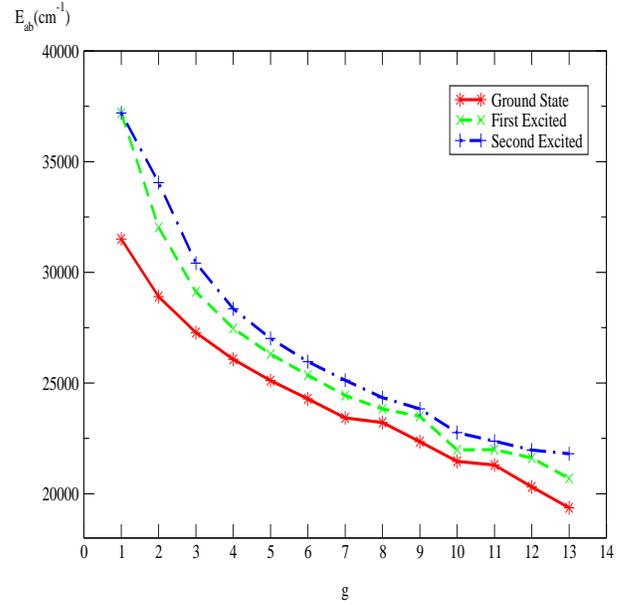}
\caption{DMRG results for the  absortion 
peaks $E_{{\rm ab}}$  (\ref{11}) in extended dendrimers (vertical
versus the generation number $g$. They
corresponds to the ground state and two excited states.}
\label{fig9}
\end{figure}

\subsubsection{Wave-Function Variational Ansatz}

Let us assume that the total number of states that we want to
keep during the truncation/renormalization process is $N_E$,
which includes the GS and $N_E-1$ excited states. These are the
targeted states and we are considering the more general case of
finding not only the GS but also a few of the low lying excited
states. Although this is not necessary for the one-exciton model
that we are considering here, however we want to show that the
method is powerful enough so as to handle more general situations.

The superblock decomposition of the lattice in turn induces a
decomposition of the wave-function of the targeted states
associated  to the blocks and the free site. Let us assume
extended and compact dendrimers with a generic value of $w$, 
then the superblock
wave-function $\Psi_l({\bf n})$ at RG-step $l$ and lattice point
${\bf n}$ is split into the following four pieces when the connectivity
is $c=3$ (\ref{13}):

\begin{equation}
\Psi_l({\bf n}) =
\begin{cases}
\sum_{\alpha=1}^{N_E} a_{\alpha} L^{\alpha}_1({\bf n};l) & \quad   {\bf n} \in B_1(l) \\
\sum_{\alpha=1}^{N_E} a_{N_E+\alpha} L^{\alpha}_2({\bf n};l) & \quad   {\bf n} \in B_2(l) \\
\sum_{\alpha=1}^{N_E} a_{2N_E+\alpha} L^{\alpha}_3({\bf n};l) & \quad   {\bf n} \in B_3(l) \\
a_{3N_E+1}            & \quad   {\bf n} = \bullet_{l} \\
\end{cases}
\label{14}
\end{equation}

\noindent where $\bullet_{l}$ denote  the free point in
Figs.\ref{fig5},\ref{fig6} connecting the blocks $B_c(l),
c=1,2,3$; $\{ L^{\alpha}_c({\bf n};l)\}_{\alpha=1}^{N_E}, c=1,2,3$
are orthonormal basis of states describing the degrees of freedom
in the wedge-blocks, i.e.,

\begin{equation}
\langle L^{\alpha}_c(l)|L^{\beta}_{c'}(l) \rangle =
\delta_{\alpha,\beta} \delta_{c,c'}
\label{15}
\end{equation}

\noindent and the free unknown coefficients
$\{a_{\alpha}\}_{\alpha=1}^{3N_E+1}$ at this $l$-stage  will be
determined later on by means of a diagonalization/truncation
process defining the renormalization. They are normalized as

\begin{equation}
\parallel {\bf a} \parallel^2 = \sum_{\alpha=1}^{3N_E+1} a_{\alpha}^2 = 1
\label{16}
\end{equation}

In a generic case of a dendrimer with $w$ wedges, there will be a number of 
blocks $c$ at each stage $l$ of the DMRG,
and a number of $cN_E+1$ variational $a$-parameters.
For compact dendrimers with $w=3$, all ansatzs are like in (\ref{14}) for
every site. However, for extended dendrimers with $w=3$ 
we have to distinguish two
types of sites, those with $c=2$ and those with $c=2$ (see Fig.~\ref{fig3})
In this case we need to use another
ansatz like in (\ref{14}) for only two blocks in (\ref{13}) 
when dealing with sites of connectivity $c=2$.

\begin{figure}[h]
\includegraphics[width=8 cm]{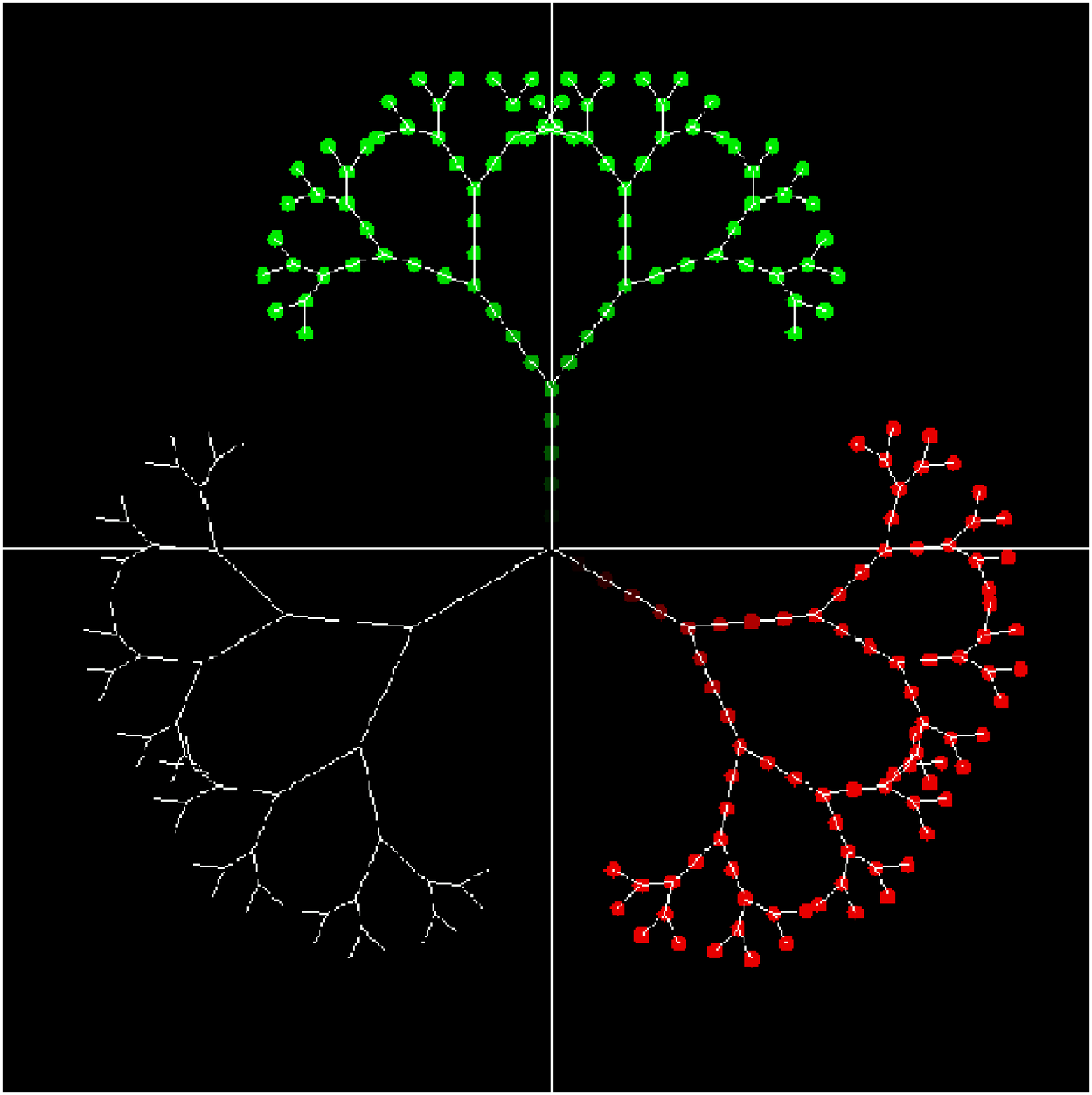}
\caption{Pictorical representation of the DMRG wavefunction for one
excited state in an extended dendrimer with generation $g=7$.
The two colors denote positive and negative values of the state coefficients
and their size are proportional to their normalized values.}
\label{fig10}
\end{figure}

\subsubsection{Superblock Hamiltonians}

The dimensionality of these SB Hamiltonians follows from the 
wavefunction in (\ref{14}), and it is much smaller
than that of the original 
Hamiltonian (\ref{12}). This makes their diagonalization something 
much less demanding than diagonalizing the whole Hamiltonian. 
The dimension of $H_{SB}$ 
depends on the number of targeted states. In the version of the
DMRG we are using here, this dimension is 
$(3N_E+1)\times(3N_E+1)$ for both families of dendrimers.
For example, when the connectivity of a given site is $c=3$ 
(where the free site (see Fig.\ref{fig5} \ref{fig6})
lies at a given DMRG-step), then the SB Hamiltonian reads as follows

\begin{equation}
H_{SB}(l) = 
\begin{pmatrix}
H_{1}(l) & 0  &  0  & v_{1}(l) \\
0    & H_{2}(l) &  0 & v_{2}(l) \\
0 & 0 & H_{3}(l) & v_{3}(l) \\
v_{1}^{\dagger}(l) & v_{2}^{\dagger}(l) & v_{3}^{\dagger}(l) & H_{\bullet_{l}\bullet_{l}}
\end{pmatrix}
\label{17}
\end{equation}

\noindent where $H_{c}(l), c=1,2,3$ are the block Hamiltonians for each
block in the DMRG decomposition of the lattice (see Fig.\ref{fig5},\ref{fig6})
with dimesions $N_E\times N_E$, namely,

\begin{equation}
H_{c}^{\alpha,\beta}  = \langle L_c^{\alpha}|H| L_c^{\beta}\rangle, \; c=1,2,3;
\label{18}
\end{equation}

\noindent while $H_{\bullet_{l}\bullet_{l}}$ 
denotes the value of the original 
Hamiltonian (\ref{12}) at the free site $\bullet_l$ and the column vectors 
$v_c(l), c=1,2,3$ with dimension $N_E\times 1$ corresponds to the interactions
between the blocks and the free site.
Their values depend on the DMRG-step $l$ and constantly updated through
the RG process. The zeroes in (\ref{17}) reflect the short-range structure
of (\ref{12}).

\begin{figure}[h]
\includegraphics[width=8 cm,  height=8 cm]{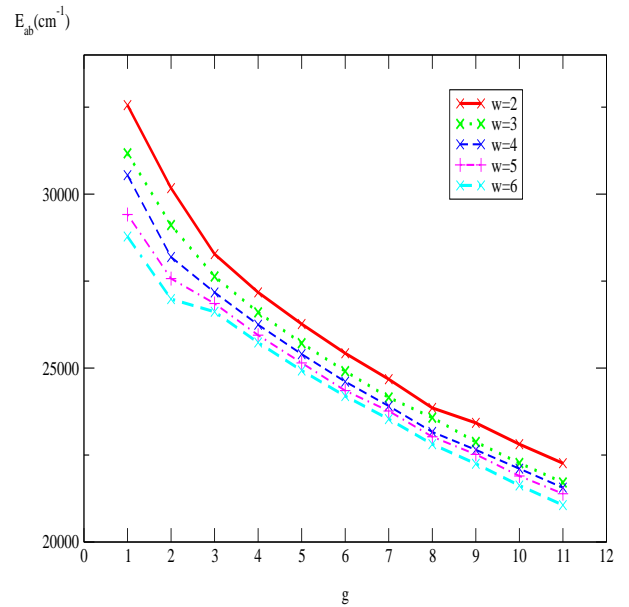}
\caption{DMRG results for the first absortion 
peak $E_{{\rm ab}}$  (\ref{11}) in extended dendrimers
versus the generation number $g$  
and with varying number of wedges $w$.}
\label{fig11}
\end{figure}

\subsubsection{Truncation of Hilbert Space and Renormalization}

The next step  
is the diagonalization of the superblock Hamiltonians
in order to obtain the 
wave functions of the $N_E$ targeted states and their energies. 
The free parameters ${\bf a}$ are then  constructed 
out of the components of the targeted wave functions. With this
information we can perform the projection of the wave functions onto
the several blocks 
as well as the renormalization of the matrix elements
of the superblock Hamiltonian.

Let us denote the variational  parameters by the set
$\{ {\bf a}_1^{i}, {\bf a}_2^{i}, {\bf a}_3^{i}, 
a_{\bullet_{l}}^{i} \}_{i=1}^{N_E}$,
where ${\bf a}_c^{i}, c=1,2,3$  are $N_E$-dimensional
vectors. Assuming that we have three blocks ($c=3$), then the truncation
of the Hilbert space is performed by the projection of the superblock
wave-function (\ref{14}) onto the blocks formed by two out of three possible
blocks plus the free site. For defineteness, let us assume that we are
renormalizing blocks $B_1(l), B_2(l)$ at a given DMRG-step $l$, 
while the remaining
block $B_3(l)$ is not directly renormalized but only updated through
the values taken from a previous DMRG-step. Then, the projection onto the
blocks $B_1(l), B_2(l)$ is given by

\begin{equation}
{\bf a}^{i} = 
\begin{pmatrix}
{\bf a}_1^{i} \\
{\bf a}_2^{i} \\
{\bf a}_3^{i} \\
a_{\bullet_{l}}^{i}\\
\end{pmatrix} \longrightarrow 
\begin{pmatrix}
{\bf a}_1^{i} \\
{\bf a}_2^{i} \\
a_{\bullet_{l}}^{i}\\
\end{pmatrix}
\label{19}
\end{equation}

\noindent The projected wave functions in the RHS of (\ref{19}) must be
orthonormalized using a Gram-Schdmit method.
They will become the new wave functions of the 
renormalized blocks $B'_1, B'_2$. More explicitly, we have

\begin{equation}
\begin{pmatrix}
{\bf a}_1^{\prime i} \\
{\bf a}_2^{\prime i} \\
a_{\bullet_{l}}^{\prime i}\\
\end{pmatrix} = \sum_j  {\cal O}^{(12)}_{i j} 
\begin{pmatrix}
{\bf a}_1^{j} \\
{\bf a}_2^{j} \\
a_{\bullet_{l}}^{j}\\
\end{pmatrix}
\label{20}
\end{equation}

\noindent where ${\cal O}^{(1)}$ is the orthonormalization matrix

\noindent The new wave functions $L_{1}^{\prime i}(l+1)$
for the renormalized  block  $B_1^{'}$
to be constructed in the next RG-step $l+1$ are given by

\begin{equation}
L_{1}^{\prime i}({\bf n};l+1) = 
\begin{cases}
\sum_\alpha  a^{\prime i}_{1, \alpha} 
L_{1}^{\alpha}({\bf n};l) & {\bf n}\in B_1(l) \\
\sum_\alpha  a^{\prime i}_{2, \alpha} 
L_{2}^{\alpha}({\bf n};l)  & {\bf n}\in B_2(l) \\
a^{\prime i}_{\bullet_{l}}  & {\bf n}=\bullet_{l}\\
\end{cases}
\label{21}
\end{equation}

\noindent After the truncation process we need to renormalize (update) the
different matrix elements of the superblock Hamiltonian (\ref{17}).
For the blocks $B_1^{'}, B_2^{'}$ we have

\begin{equation}
H^{\prime}_1(l+1)  
= 
A^{\prime \dagger}
\begin{pmatrix}
H_{1}(l) & 0 & v_{1}(l)  \\
0 & H_{1}(l) & v_{2}(l)  \\
v_{1}^{\dagger}(l) & v_{2}^{\dagger}(l) & H_{\bullet_{l},\bullet_{l}}\\
\end{pmatrix}
A^{\prime}
\label{22}
\end{equation}

\begin{equation}
A^{\prime} = 
\begin{pmatrix}
{\bf a}_1^{\prime 1} & \ldots & {\bf a}_1^{\prime N_E} \\
{\bf a}_2^{\prime 1} & \ldots & {\bf a}_2^{\prime N_E} \\
a_{\bullet_{l}}^{\prime 1} & \ldots & a_{\bullet_{l}}^{\prime N_E}\\
\end{pmatrix}
\label{23}
\end{equation}

These considerations for the renormalization of blocks $B_1(l), B_2(l)$ are
generic and they equally apply to the other pairs of blocks.
Moreover, when the free site is placed at a lattice site with connectivity
$c=2$, then the renormalization is simplified for then there are only 2
blocks and the RG process is similar to a one-dimensional DMRG step,
with a lattice decomposition of the type $B\bullet B$.

\subsubsection{Sweeping Dendrimers}

This part of the finite-size DMRG method corresponds to moving the free
site in Fig.~\ref{fig6} throughout the lattice at each step of the RG
process and updating the content of the blocks in the DMRG decomposition
of the lattice following the previous renormalization prescriptions. 
The sweeping process is responsible for the convergence
of the method: the sweeping continues until the energy values of the
intermediate diagonalizations achieve a desired prescribed precision.

During a generic DMRG sweep, the free site starts at the focal point and then 
moves first through the sites in wedge $w=1$, next the sites of $w=2$ are
visited and so on and so forth with the rest of the wedges until the free
site returns to the focal point. The sweeping process needs a first input
called warm-up in order to start up. In the case of dendrimers we find 
useful to do it with a simple block renormalization group method (BRG).
This means that the dendrimer lattice is decomposed in blocks.
The blocks in this BRG are formed with 3 sites,
one of these sites is the father and the other two are children.
The first step of this BRG starts with a blocking for the outermost sites
located at the surface of the dendrimer, namely, the last and next-to-last
generations. During the following BRG-steps, the dendrimer lattice is 
reconstructed from the outside to the focal point located at the core.
At this moment, the warm-up process is completed.
Interestingly enough, this form of doing the warm-up in the DMRG for dendrimres
is similar to one of the actual methods of chemical 
synthesization: the convergent method \cite{frechet}.

\subsection{Numerical Results}

Let us start presenting our numerical results for the family of
extended dendrimers. Our main purpose is to show the capabilities
of the DMRG method when dealing with one-exciton problems in the 
presence of disorder 
such the case of the  Harigaya model that we have taken as our
model Hamiltonian (\ref{8},\ref{9}).

We have targeted a number $N_E=3$ of states that include the ground state
plus two excited states. The convergence criterion used is to sweep until
a precision (relative error) of $10^{-10}$ is achive in all the eigenvalues.
We find that the degree of convergence with our DMRG is very fast. 
For instance, the number of sweeps needed to achieve this accuracy is always
lower than 4. 
The number of samples is also $N_S=10000$ as in 
\cite{harigaya1,harigaya2,harigaya3}. 
The corresponding  CPU times are also very small as compared
with exact diagonalization times. For example,  for the sampling considered
here a typical time for the dendrimers with higher generation number is 
around 900 s \cite{pc}.
We start with the number of wedges $w=3$.

In Fig.~\ref{fig7} we present the results for the first absortion 
peak $E_{{\rm ab}}$  (\ref{11}) in extended dendrimers versus the
generation number $g$ until a value of 7. The energies are measured
in ${\rm cm}^{-1}$.
Here we are doing two types of comparisons. Firstly, we are comparing our
DMRG results with the exact diagonalization results (ED) of Harigaya
up to 5 generations. We clearly see that there is an overlapping
of both results and furthermore, we can go easily to higher generations.
Secondly, we must note in passing that there is 
qualitative agreement between the
experimental results of Kopelman et al. \cite{kopelman} and those of 
Harigaya's, although a better quantitative matching would be desireable.

We have been able to reach a generation number of $g=13$ without using
big computer facilities \cite{pc} as shown in Fig.~\ref{fig8}. The vertical
lines at each result in this figure are error
bars due to statistical fluctuations in the sampling. 
It is apparent from these results that
there is a decreasing of the absortion peak as the generation number increases,
in qualitative agreement with the experimental results of 
Kopelman et al. \cite{kopelman}.
Although in  \cite{kopelman} only extended and compact dendrimers up to
$g=5$ were considered, it has been possible experimentally to reach 
higher generation numbers in a variety of dendrimers of different 
compositions. In our case, we are also interested in studying theoretically
the behaviour with $g$ of physical properties in dendrimers. 
As it happens, due to the exponential growth of the number of sites with
$g$, the dendrimer tends to reach an upper generation limit because the 
surface groups become densely packed. This is known as the ``de Gennes 
dense packing''. For some dendrimers, this limit has been found around
$g=10$ to $g=12$.

As a result of having this many values
we can make a polynomial fit to stimate the degree of this slowing down.
We find that the following scaling law with the generation number $g$

\begin{equation}
E_{{\rm ab}}^{{\rm e}} = 32000 - C_e  g^{\alpha_e} \ {\rm cm}^{-1},
\label{33}
\end{equation}

\noindent fits  well the numerical data with the values 
$C_e=2020 \pm 200 {\rm cm}^{-1}$ 
and the scaling exponent $\alpha_e=0.71 \pm 0.04$. The continuous line in
Fig.~\ref{fig8} gives the fitting in (\ref{33}).

As we are interested in the performance of the DMRG method when dealing
with excitons, we also plot two byproducts of our numerical calculations.
On one hand, in Fig.~\ref{fig9} we present the results for the 
absortion  peaks $E_{{\rm ab}}$  (\ref{11}) corresponding not only to
the ground state, but also to the lowest two excited states. They are
computed simultaneously when targeting $N_E=3$ states. We see that all 
of these states have decreasing energies as the generation number increase.
Within the approximations of the Harigaya model, this is an indication
that these lowest lying states are delocalized.
On the other hand, in Fig.~\ref{fig10} we show a plot of the wavefunction
corresponding to an excited state for a generation number of $g=7$. 
This is because the DMRG not only gives
energies but also computes the corresponding states. In this pictorical
view of the excited state in the extended dendrimer, each circle has a size
proportional to the value of the coefficient of the normalized wavefunction
and the two colors corresponds to positive and negative values. Notice that
in one of the wedges the state has vanishing contributions.

So far we have considered the number of wedges fixed to $w=3$, following
the experimental studies in \cite{kopelman} and the computations in
\cite{harigaya1,harigaya2,harigaya3}. However, as for several types of
dendrimers the number of wedges achieved experimentally can be different 
than this number \cite{lobulos0,lobulos1,lobulos2}, we have also applied
our DMRG studies to deal with extended dendrimers with variable number
of wedges ranging from $w=2$ to $w=6$ as shown in Fig.~\ref{fig11}.
In this figure we present the first absortion peaks $E_{{\rm ab}}$  (\ref{11})
as a function of $g$ and with varying number $w$. It is clearly seen
that the effect of increasing $w$ is to lower the energy peaks for any
generation number (there are no level crossings). For a fixed value of
$w$, the behaviour of the peaks are all qualitatively similar.

\begin{figure}[h]
\includegraphics[width=8 cm, height=8 cm]{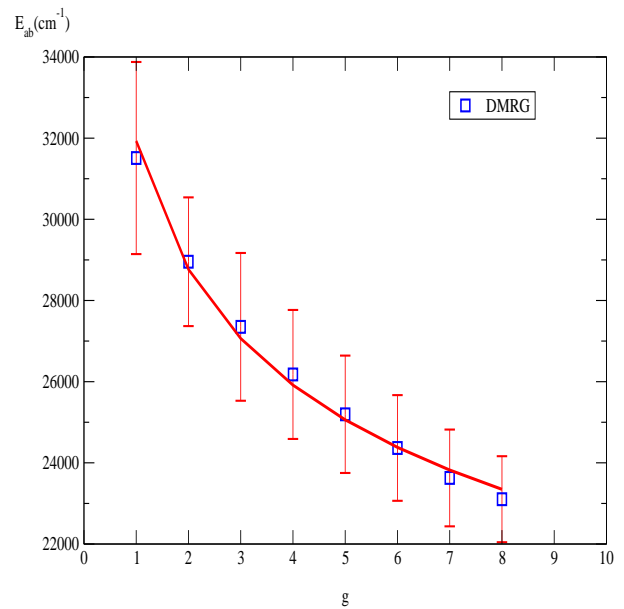}
\caption{Results for the first absortion 
peak $E_{{\rm ab}}$  (\ref{11}) using DMRG in 
compact dendrimers versus the generation number $g$\cite{sampling}.}
\label{fig12}
\end{figure}

For the sake of completeness, we have also performed a DMRG study in the
family of compact dendrimers with a fixed number of wedges $w=3$ using
the Harigaya model (\ref{8}),(\ref{9}). We plot the first absortion 
peak $E_{{\rm ab}}$  (\ref{11}) as a function of $g$ up to 8, with
the corresponding error bars. From this figure it is shown that the energy
peaks also decrease with increasing sizes for this family of dendrimers
and thus the Harigaya model is not able to predict the correct experimental
data in \cite{kopelman}.
We can make a polynomial fit to stimate again the degree of this decay.
We find that the following scaling law 

\begin{equation}
E_{{\rm ab}}^{{\rm c}} =  32000 - C_c  g^{\alpha_c} \ {\rm cm}^{-1},
\label{34}
\end{equation}

\noindent fits very  well the numerical data with the values 
$C_c=3190 \pm 120 {\rm cm}^{-1}$ 
and the scaling exponent $\alpha_c=0.54 \pm 0.02$. The continuous line in
Fig.~\ref{fig12} gives the fitting in (\ref{34}). Thus, the only difference 
that we can capture between both families of dendrimers is the fact that
the scaling exponent $\alpha_e$  in the extended dendrimers is bigger
than $\alpha_c$  for the compact ones.

\section{Conclusions and Prospects}
\label{sec4:level1}

The field of dendrimers has become very active in Chemistry 
during the last decade due to their many possible applications 
explained in the introduction 
and we believe that there are open physical problems and new physics
so as to make also an entrance in condensed matter also.

In this work we have been interested in the one-exciton processes 
pertaining to the application of extended dendrimers as 
light-harvesting molecules, i.e., molecular antennas.
Our main contribution is to introduce the DMRG method as a very appropriate
computational tool for one-exciton approximations with off-diagonal
disorder. This is because the renormalization/truncation
process in a DMRG calculation is more efficient than the exact diagonalization
method if we are only interested in the low-lying states of the model
Hamiltonian. The gain of the DMRG method is more pronounced when a large
number of diagonalizations are needed such the large samplings 
typical of disorder problems with excitons.

We have shown the potentialities of the DMRG method in the field of 
excitons with disorder by computing a variety of dendrimer configurations
with a variable number of generations $g$ and wedges $w$ (\ref{1}).
In all our computations we reproduce the exact results for low generation
number and further, we can increase $g$ without much computational effort
to numbers that become unreachable for exact diagonalizations techniques (ED).
Thus we propose to substitute these ED methods by the DMRG in situations like
two-dimensional square lattices where the computation of one-exciton
results with disorder and long-range interactions becomes very demanding
for ED methods. For instance, we have computed lattices with a number of
36823 sites in the case of extended dendrimers, just to give an example
of the sizes that can be reached with DMRG.

One of the open questions in the field of extended/compact dendrimers is
the explanation of the scenario in (\ref{2}) regarding the closing or
not of the gap in the absortion spectrum
using a more elaborate model
Hamiltonian than the oversimplified Harigaya model 
\cite{harigaya1,harigaya2,harigaya3}.
It is plausible to think that a truly many-body Hamiltonian would be 
necessary in any attemp to address this issue.
If we concentrate only in the electronic properties of these compounds,
one possible candidate to describe the physics of multi-excitonic processes
is a PPP (Pariser-Parr-Pople) Hamiltonian \cite{review-polymers,rva1,rva2}. 
This type of Hamiltonian has been used with success in conjugated 
polymers of linear types and we believe that it could also be of use for
conjugated polymer of dendrimer type. 
Another simpler candidate for a many-body Hamiltonian description of the
gap spectrum in extended/compact dendrimers is the Frenkel Hamiltonian 
introduced in (\ref{5}) that is more manageable for a first try than the 
PPP Hamiltonian.
These types of studies are  left for future work.

\begin{acknowledgments}

We would like to thank K. Harigaya for reading a draft version of this
work.  This work has been partially supported by the Spanish grant PB98-0685.

\end{acknowledgments}


\begin{thebibliography}{99}
\bibitem{dend} Dendrimer = dendrite + dimer.
From the Greek word dendra = tree.
Dendrites are the terminations of brain cells (neurons),
while dimers are bonds (singlets).
\bibitem{xy} The excitonic operators $b_i$ are in fact hard-core bosons.
The local states $|{\rm vac}\rangle_i$, $|{\rm exc}\rangle_i$ are in one-to-one
correspondence with spin states $|-\rangle_i$, $|+\rangle_i$ and likewise, the
operators $b_i$, $b^{\dagger}_i$ and  $ b^{\dagger}_i  b_i$ can be represented
by Pauli matrices $\sigma^-_i$, $\sigma^+_i$ and $(\sigma^z_i+1)/2$,
respectively. Thus, the Hamiltonian (\ref{5}) becomes the XY model with
transverse couplings given by $J_{i,j}$ and external magnetic field given
by $E_i$.
\bibitem{sampling} The error bars for $g=12,13$ are  bigger 
in these generations just because the sampling number is reduced,
namely, $N_S=100$.
\bibitem{pc} We have used a personal computer with a Pentium III at 450 MHZ.
\end{thebibliography}
\end{document}